\documentclass[preprint,showpacs,preprintnumbers,eqsecnum,floatfix]{revtex4}
\usepackage{graphicx}
\usepackage{dcolumn}
\usepackage{bm}

\font\scripti=cmmi7
\font\scriptscripti=cmmi5
\def\sib#1{\setbox0 = \hbox{\scripti #1}
  \kern-.02em\copy0\kern-\wd0
  \kern.04em\box0} 
\def\ssib#1{\setbox0 = \hbox{\scriptscripti #1}
  \kern-.02em\copy0\kern-\wd0
  \kern.04em\box0} 
\font\tenib=cmmib10 
\skewchar\tenib='177 \skewchar\tenib='177 \skewchar\tenib='177
\def\pbold#1{\setbox0 = \hbox{$ #1 $}
  \kern-.022em\copy0\kern-\wd0
  \kern.011em\copy0\kern-\wd0
  \kern.011em\copy0\kern-\wd0
  \kern.011em\copy0\kern-\wd0
  \kern.011em\box0} 

\usepackage{graphicx}
\usepackage{dcolumn}
\usepackage{bm}

\def\s{\sigma}

\def\up{\uparrow}
\def\dwn{\downarrow}

\def\lesssim{\ \raise.3ex\hbox{$<$}\kern-0.8em\lower.7ex\hbox{$\sim$}\ }
\def\gesim{\ \raise.3ex\hbox{$>$}\kern-0.8em\lower.7ex\hbox{$\sim$}\ }

\begin{document}
\title{Spin susceptibility and effects of a harmonic trap in the BCS-BEC crossover regime of an ultracold Fermi gas}
\author{Hiroyuki Tajima$^1$, Ryo Hanai$^2$, and Yoji Ohashi$^3$}
\affiliation{$^1$RIKEN Nishina Center, Wako 351-0198, Japan}
\affiliation{$^2$Department of Physics, Osaka University, Toyonaka 560-0043, Japan}
\affiliation{$^3$Department of Physics, Keio University, 3-14-1 Hiyoshi, Kohoku-ku, Yokohama 223-8522, Japan}
\date{\today}
\begin{abstract}
We theoretically investigate magnetic properties of a trapped ultracold Fermi gas. Including pairing fluctuations within the framework of an extended $T$-matrix approximation (ETMA), as well as effects of a harmonic trap in the local density approximation (LDA), we calculate the local spin susceptibility $\chi_{\rm t}(r,T)$ in the BCS (Bardeen-Cooper-Schrieffer)-BEC (Bose-Einstein condensation) crossover region. We show that pairing fluctuations cause non-monotonic temperature dependence of $\chi_{\rm t}(r,T)$. Although this behavior looks similar to the spin-gap phenomenon associated with pairing fluctuations in a {\it uniform} Fermi gas, the trapped case is found to also be influenced by the temperature-dependent density profile, in addition to pairing fluctuations. We demonstrate how to remove this extrinsic effect from $\chi_{\rm t}(r,T)$, to study the interesting spin-gap phenomenon purely originating from pairing fluctuations. Since experiments in cold atom physics are always done in a trap, our results would be useful for the assessment of preformed pair scenario, from the viewpoint of spin-gap phenomenon. 
\end{abstract}
\maketitle
\par
\section{Introduction}
\par
Since the realization of superfluid $^{40}$K \cite{Regal} and $^6$Li \cite{Zwierlein,Kinast,Bartenstein} Fermi gases, strong-coupling properties in the BCS (Bardeen-Cooper-Schrieffer)-BEC (Bose-Einstein condensation) crossover region have attracted much attention in this field \cite{Holland,Ohashi2,Levin2005,Giorgini,Bloch,Chin,Zwerger}. In this regime, the system properties are dominated by strong pairing fluctuations, that are physically described as repeating the formation and dissociation of preformed (Cooper) pairs\cite{Nozieres,SadeMelo,Randeria,Randeria1992}. Thus, ultracold Fermi gases are expected to provide a useful testing ground for the assessment of the so-called preformed pair scenario, which has been proposed as a possible mechanism of the pseudogap observed in the underdoped regime of high-$T_{\rm c}$ cuprates \cite{Randeria1992,Singer,Janko,Renner,Yanase,Rohe,Perali,Lee2,Fischer,Varma} (where a gap-like structure appears in the density of states (DOS) even above the superconducting phase transition temperature $T_{\rm c}$). Although the origin of this anomaly is still unclear in high-$T_{\rm c}$ cuprates because of the complexity of this electron system \cite{Randeria1992,Singer,Janko,Renner,Yanase,Rohe,Perali,Lee2,Fischer,Varma,Pines,Kampf,Chackravarty}, if it is observed in an ultracold Fermi gas, the origin must be strong pairing fluctuations, or the formation of preformed Cooper pairs \cite{Tsuchiya,Watanabe,Mueller,Magierski,Su}. Although this observation would not immediately clarify the pseudogap phenomenon in high-$T_{\rm c}$ cuprates, one may regard it as an evidence for the validity of the preformed pair scenario, at least, in the presence of strong pairing fluctuations. 
\par
At present, the pseudogap has not been observed in an ultracold Fermi gas yet, because of the difficulty of the direct observation of DOS in this field. Although a photoemission-type experiment supports the preformed pair scenario\cite{Stewart,Gaebler,Sagi2}, thermodynamic measurements\cite{Nascimbene2,Nascimbene3,Nascimbene} report the Fermi liquid-like behavior of the system with no pseudogap. Thus, further studies are necessary to resolve this controversial situation.
\par
Recently, we have theoretically pointed out \cite{Tajima,Tajima2} that the spin-gap may be an alternative key phenomenon to assess the preformed pair scenario in an ultracold Fermi gas. This magnetic phenomenon is characterized by the anomalous suppression of spin susceptibility in the normal state near the superfluid phase transition temperature. This many-body phenomenon has been observed in high-$T_{\rm c}$ cuprates \cite{Takigawa,Yoshinari,Takigawa2,Bobroff,Ogata}, although the origin is still in controversial. In the preformed pair scenario, the pseudogap and spin-gap are understood as different aspects of the same pairing phenomenon. That is, while the former is explained from the viewpoint of ``binding energy" of preformed pairs, the latter is understood as a result of the formation of {\it spin-singlet} preformed pairs. Indeed, it has theoretically been shown \cite{Tajima} that the pseudogap temperature (below which a dip structure appears in DOS) is very close to the spin-gap temperature (below which the spin susceptibility is anomalously suppressed) in the BCS-unitary regime. Recently, the spin susceptibility has become observable in cold Fermi gas physics \cite{Sanner,Sommer,Meineke,Lee}, and theoretical analyses on the observed spin susceptibility have been started \cite{Tajima,Tajima2,Palestini,Mink,Enss}. Thus, this alternative approach seems promising in the current stage of cold Fermi gas physics.
\par
In this paper, we extend our previous work \cite{Tajima,Tajima2} for the spin susceptibility in a uniform Fermi gas to include effects of a harmonic trap. This extension is really important, because experiments are always done in a trap potential. Thus, it is a crucial issue how spatially inhomogeneous pairing fluctuations affect spin susceptibility. In addition, to assess the preformed pair scenario without any ambiguity, we need to know spin susceptibility in a {\it uniform} Fermi gas, from observed data in a {\it trapped} Fermi gas. Regarding this, the pseudogap case is simpler, because, once the local density of states $\rho({\bm r},\omega)$ becomes observable in the future, the observed dip structure in $\rho({\bm r},\omega)$ around $\omega=0$ can immediately be interpreted as the pseudogap in the uniform case with the uniform density $n$ being equal to the local density $n({\bm r},T)$ at the observed spatial position ${\bm r}$. On the other hand, since the spin-gap appears in the {\it temperature dependence} of the spin susceptibility, it is sensitive to the temperature-dependence of the density profile $n({\bm r},T)$. To examine the preformed pair scenario proposed in the uniform system, we need to remove the latter extrinsic effect from the observed temperature dependence of the spin susceptibility in a trap. 
\par
For our purpose, we include effects of a harmonic trap in the local density approximation (LDA) \cite{Ohashi6,Perali3,Haussmann3,Tsuchiya2,Tsuchiya3,Watanabe}. Pairing fluctuations are taken into account within the framework of an extended $T$-matrix approximation (ETMA) \cite{Tajima,Tajima2,Kashimura,Hanai,Kashimura2,Tajima3}. Using this combined theory, we calculate the local spin susceptibility $\chi_{\rm t}(r,T)$, as well as the spatially averaged one $X_{\rm t}(T)$, in the whole BCS-BEC crossover region. We demonstrate how we can map $\chi_{\rm t}(r,T)$ onto the spin susceptibility $\chi_{\rm u}(T)$ in a {\it uniform} Fermi gas. We also compare our results with the recent experiment in a $^6$Li Fermi gas \cite{Sanner}. 
\par
This paper is organized as follows. In Sec. II, we present our combined extended $T$-matrix approximation (ETMA) with the local density approximation (LDA). In Sec. III, we show our numerical results on the local spin susceptibility $\chi_{\rm t}(r,T)$ in the BCS-BEC crossover region of a trapped Fermi gas. Here, we also explain how to relate $\chi_{\rm t}(r,T)$ to $\chi_{\rm u}(T)$ in a uniform Fermi gas, to examine the spin-gap phenomenon purely originating from pairing fluctuations. In Sec. IV, we consider the trap-averaged spin susceptibility $X_{\rm t}(T)$, to compare our results with the recent experiment on a $^6$Li Fermi gas \cite{Sanner}. Throughout this paper, we set $\hbar=k_{\rm B}=1$, for simplicity.
\par
\section{Formulation}
\par
To explain our formalism, we start from a uniform superfluid Fermi gas. Effects of a harmonic trap will be included later. In the two-component Nambu representation \cite{Fukushima,Schrieffer,Ohashi4,Watanabe}, our model Hamiltonian is given by
\begin{equation}
H=\sum_{\bm p}
{\hat \Psi}_{\bm p}^\dagger
\left[
\xi_{\bm p}\tau_3-h-\Delta\tau_1
\right]
{\hat \Psi}_{\bm p }
-{U \over 4}\sum_{\bm q}
\left[
\rho_{1,{\bm q}}\rho_{1,-{\bm q}}+
\rho_{2,{\bm q}}\rho_{2,-{\bm q}}
\right].
\label{eq1}
\end{equation}
Here,
\begin{eqnarray}
{\hat \Psi}_{\bm p}=
\left(
\begin{array}{c}
c_{{\bm p},\uparrow} \\
c_{-{\bm p},\downarrow}^\dagger
\end{array}
\right)
\label{eq1b}
\end{eqnarray}
is the two-component Nambu field, where $c_{{\bm p},\sigma}^\dagger$ is the creation operator of a Fermi atom with pseudospin $\sigma=\uparrow,\downarrow$, describing two atomic hyperfine states. $\xi_{\bm p}=\varepsilon_{\bm p}-\mu={\bm p}^2/(2m)-\mu$ is the kinetic energy of a Fermi atom, measured from the Fermi chemical potential $\mu$, where $m$ is an atomic mass. Although we consider the population-balanced case, the model Hamiltonian in Eq. (\ref{eq1}) involves an infinitesimally small fictious magnetic field $h$, in order to calculate the spin susceptibility later.  $\tau_i$ ($i=1,2,3$) are Pauli matrices acting on particle-hole space. In Eq. (\ref{eq1}), the superfluid order parameter $\Delta$ is taken to be parallel to the $\tau_1$-component, without loss of generality. In this choice, the generalized density operators,
\begin{equation}
\rho_{i,{\bm q}}=
\sum_{\bm p}{\hat \Psi}_{{\bm p}+{\bm q}/2}^\dagger\tau_i
{\hat \Psi}_{{\bm p}-{\bm q}/2}~~(i=1,2),
\label{eq1c}
\end{equation}
physically describe amplitude fluctuations $(i=1)$ and phase fluctuations $(i=2)$ of the superfluid order parameter $\Delta$ \cite{Fukushima,Ohashi4,Watanabe}. We briefly note that the ordinary contact-type $s$-wave pairing interaction is described by the sum of amplitude-amplitude ($\rho_1\rho_1$) and phase-phase ($\rho_2\rho_2$) interactions in Eq. (\ref{eq1}). 
\par
As usual, we measure the interaction strength in terms of the $s$-wave scattering length $a_s$, which is related to the bare coupling constant $-U~(<0)$ as,
\begin{equation}
{4\pi a_s \over m}=-
{U \over \displaystyle 1-U\sum_{\bm p}{1 \over 2\varepsilon_{\bm p}}}.
\label{eq2}
\end{equation}
In this scale, the weak-coupling BCS regime and strong-coupling BEC regime are conveniently characterized as $(k_{\rm F}a_s)^{-1}\lesssim -1$ and $(k_{\rm F}a_s)^{-1}\gesim 1$, respectively. The region between the two is called the BCS-BEC crossover region. 
\par
\begin{figure}[t]
\begin{center}
\includegraphics[width=8cm]{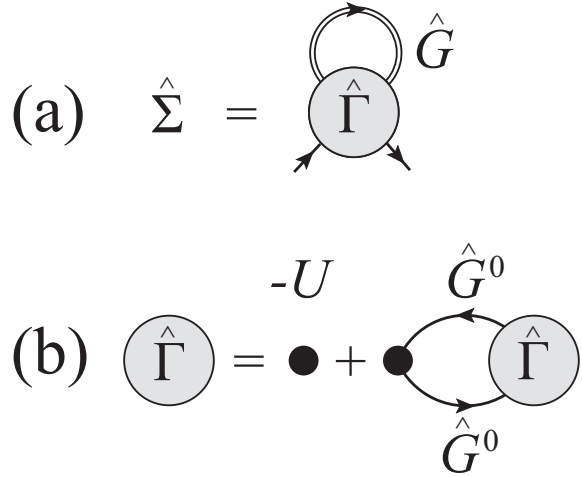}
\end{center}
\caption{(a) Self-energy correction ${\hat \Sigma}$ in combined ETMA with LDA (LDA-ETMA). The $2\times 2$-matrix particle-particle scattering vertex ${\hat \Gamma}=\{\Gamma\}^{j,j'}$ is given in (b). In this figure, the double and single solid lines denote the LDA-ETMA dressed Green's function ${\hat G}$ in Eq. (\ref{eq4}), and the bare one ${\hat G}^{0}$ in Eq. (\ref{eq4b}), respectively. The filled circle is a pairing interaction $-U$.}
\label{fig1}
\end{figure}
\par
Now, we include effects of a harmonic trap. In the local density approximation (LDA)\cite{Ohashi6,Perali3,Haussmann3,Tsuchiya2,Tsuchiya3,Watanabe}, this extension is achieved by simply replacing the Fermi chemical potential $\mu$ by the LDA one, $\mu(r)=\mu-V(r)$, where
\begin{equation}
V(r)={1 \over 2}m\Omega_{\rm tr}^2 r^2
\label{eq3}
\end{equation}
is a harmonic potential with a trap frequency $\Omega_{\rm tr}$, with $r$ being the radial position, measured from the trap center. The $2\times 2$-matrix single-particle thermal Green's function in LDA has the form, 
\begin{equation}
{\hat G}_{\bm p}(i\omega_n,r)=
{1 \over 
(i\omega_n+h)-\xi_{\bm p}(r)\tau_3+\Delta(r) \tau_1-\hat{\Sigma}_{\bm p}
(i\omega_n,r)},
\label{eq4}
\end{equation}
where $\omega_n$ is the fermion Matsubara frequency, $\xi_{\bm p}(r)=\varepsilon_{\bm p}-\mu(r)$, and $\Delta(r)$ is the LDA position-dependent superfluid order parameter. The $2\times 2$-matrix self-energy ${\hat \Sigma}_{\bm p}(i\omega_n,r)$ describes strong-coupling corrections to single-particle excitations. Within the framework of the combined extended $T$-matrix approximation (ETMA) with LDA (LDA-ETMA), it is diagrammatically described as Fig. \ref{fig1}, which gives
\begin{equation}
{\hat \Sigma}_{\bm p}(i\omega_n,r)=-T
\sum_{{\bm q},\nu_n}\sum_{j,j'=\pm}\Gamma^{j,j'}_{\bm q}(i\nu_n,r)
\tau_j {\hat G}_{{\bm p}+{\bm q}}(i\omega_n+i\nu_n,r)\tau_{j'}.
\label{eq5}
\end{equation}
Here, $\nu_n$ is the boson Matsubara frequency, $\tau_{\pm}=(\tau_1+i\tau_2)/2$, and 
\begin{eqnarray}
\left(\begin{array}{cc}
\Gamma^{-+}_{\bm{q}}(i\nu_n,r) & \Gamma^{--}_{\bm{q}}(i\nu_n,r) \\
\Gamma^{++}_{\bm{q}}(i\nu_n,r) & \Gamma^{+-}_{\bm{q}}(i\nu_n,r) 
\end{array}\right) 
=-U\left[1+U
\left
(\begin{array}{cc}
\Pi^{-+}_{\bm{q}}(i\nu_n,r) & \Pi^{--}_{\bm{q}}(i\nu_n,r) \\
\Pi^{++}_{\bm{q}}(i\nu_n,r) & \Pi^{+-}_{\bm{q}}(i\nu_n,r) 
\end{array}
\right)\right]^{-1}
\label{eq6}
\end{eqnarray}
is the $2\times 2$-matrix particle-particle scattering vertex, describing fluctuations in the Cooper channel. In Eq. (\ref{eq6}),
\begin{equation}
\Pi^{j,j'}_{\bm q}(i\nu_n,r)=
T\sum_{{\bm p},i\omega_n}{\rm Tr}
\left[
\tau_j\hat{G}^0_{{\bm p}+{\bm q}}(i\omega_n+i\nu_n,r)
\tau_{j'}\hat{G}^0_{\bm p}(i\omega_n,r)
\right],
\label{eq7}
\end{equation}
is the lowest-order pair correlation function, where 
\begin{equation}
\hat{G}^{0}_{\bm p}(i\omega_n,r)
={1 \over i\omega_n-\xi_{\bm p}(r)\tau_3+\Delta(r)\tau_1}
\label{eq4b}
\end{equation}
is the $2\times 2$-matrix mean-field BCS single-particle thermal Green's function.
\par
An advantage of ETMA is that one can obtain the expected {\it positive} spin susceptibility in the whole BCS-BEC crossover region\cite{Kashimura}. The ordinary (non-selfconsistent) $T$-matrix approximation (TMA), as well as the strong-coupling theory developed by Nozi\`eres and Schmitt-Rink (NSR), are known to unphysically give {\it negative} spin susceptibility in the crossover region, because of unsatisfactory treatment of strong-coupling corrections to spin-vertex and single-particle density of states, respectively \cite{Kashimura}.
\par
We calculate the local spin susceptibility $\chi_{\rm t}(r,T)$ from
\begin{equation}
\chi_{\rm t}(r,T)=
\lim_{h\rightarrow 0}{n_\uparrow(r,T)-n_\downarrow(r,T) \over h}.
\label{eq11}
\end{equation}
Here, $n_\sigma(r,T)$ is the density profile of $\sigma$-spin atoms, which is calculated from LDA-ETMA dressed Green's function in Eq. (\ref{eq4}) as,
\begin{eqnarray}
\begin{array}{l}
\displaystyle
n_{\up}(r,T)=T\sum_{{\bm p},i\omega_n}G^{11}_{\bm p }(i\omega_n,r), 
\\
\displaystyle
n_{\dwn}(r,T)=\sum_{\bm{p}}1-
T\sum_{{\bm p},i\omega_n}G^{22}_{\bm p}(i\omega_n,r).
\end{array}
\label{eq10}
\end{eqnarray}
In this paper, we numerically evaluate Eq. (\ref{eq11}), by setting $h/\varepsilon_{\rm F}^{\rm t}=0.01$, where $\varepsilon_{\rm F}^{\rm t}$ is the Fermi energy of a trapped free Fermi gas. We have numerically confirmed that the difference $n_\uparrow(r)-n_\downarrow(r)$ is proportional to $h$, when $h/\varepsilon_{\rm F}^{\rm t}=O(10^{-2})$. 
\par
In this paper, we also consider the spatially averaged (or total) spin susceptibility,
\begin{equation}
X_{\rm t}(T)=
\int d{\bm r}\chi_{\rm t}(r,T)=\lim_{h\rightarrow0}\frac{N_{\up}-N_{\dwn}}{h},
\label{eq12}
\end{equation}
where $N_\sigma$ is the number of $\sigma$-spin atoms.
\par
In calculating Eqs. (\ref{eq11}) and (\ref{eq12}), we note that effects of fictitious field $h$ on the superfluid order parameter $\Delta(r)$ and the chemical potential $\mu$ are $O(h^2)$. For example, the gap equation, which is obtained from the condition for the gapless Goldstone mode (${\rm det}[{\hat \Gamma}_{{\bm q}=0}(\nu_m=0,r)]=0$, where ${\hat \Gamma}=\{\Gamma^{j,j'}\}$), has the form, in the presence of $h$, 
\begin{equation}
1=-{4\pi a_s \over m}
\sum_{\bm p}
\left[
{1 \over 4E_{\bm p}(r)}
\left[
\tanh{E_{\bm p}(r)+h \over 2T}
+
\tanh{E_{\bm p}(r)-h \over 2T}
\right]
-{1 \over 2\varepsilon_{\bm p}}
\right],
\label{eq11b}
\end{equation}
where $E_{\bm p}(r)=\sqrt{\xi_{\bm p}^2(r)+\Delta^2(r)}$ describes local Bogoliubov single-particle excitations in LDA. The right-hand side of Eq. (\ref{eq11b}) is clearly an even function of $h$, indicating the even function of $\Delta(r)$ in terms of $h$. Because of this, we can safely ignore $h$ in determining $\Delta(r)$ and $\mu$ for our purpose. The gap equation (\ref{eq11b}) is then simplified as ($h=0$),
\begin{equation}
1=-{4\pi a_s \over m}
\sum_{\bm p}
\left[{1 \over 2E_{\bm p}(r)}\tanh{E_{\bm p}(r) \over 2T}-
{1 \over 2\varepsilon_{\bm p}}
\right].
\label{eq8}
\end{equation}
We solve Eq. (\ref{eq8}), together with the equation for the total number $N$ of Fermi atoms,
\begin{equation}
N=\sum_\sigma N_\sigma=\sum_\sigma\int d^3{\bm r}n_\sigma(r,T)_{h=0},
\label{eq9}
\end{equation}
to self-consistently determine $\Delta(r)$ and $\mu$.
\par
Although the LDA gap equation (\ref{eq8}) gives position-dependent superfluid phase transition temperature $T^{\rm t}_{\rm c}(r)$, it is an artifact of this approximation.  The superfluid order parameter should become finite everywhere in a gas cloud below the superfluid phase transition $T^{\rm t}_{\rm c}$ of the system. In this sense, $T^{\rm t}_{\rm c}(r)$ should physically be regarded as a characteristic temperature below which the superfluid order parameter at $r$ becomes large. In LDA, the superfluid phase transition temperature $T_{\rm c}^{\rm t}$ is determined from the $T_{\rm c}^{\rm t}(r)$-equation at $r=0$,
\begin{equation}
1=-{4\pi a_s \over m}
\sum_{\bm p}
\left[{1 \over 2\xi_{\bm p}}\tanh{\xi_{\bm p} \over 2T^{\rm t}_{\rm c}}-
{1 \over 2\varepsilon_{\bm p}}
\right].
\label{eq8b}
\end{equation}
Above $T^{\rm t}_{\rm c}$, as well as in the spatial region with vanishing superfluid order parameter $\Delta(r)=0$ even below $T^{\rm t}_{\rm c}$ (Note that $T_{\rm c}^{\rm t}(r)\le T_{\rm c}^{\rm t}$.), we only solve the number equation (\ref{eq9}), to determine $\mu$.
\par
\begin{figure}[t]
\begin{center}
\includegraphics[width=8cm]{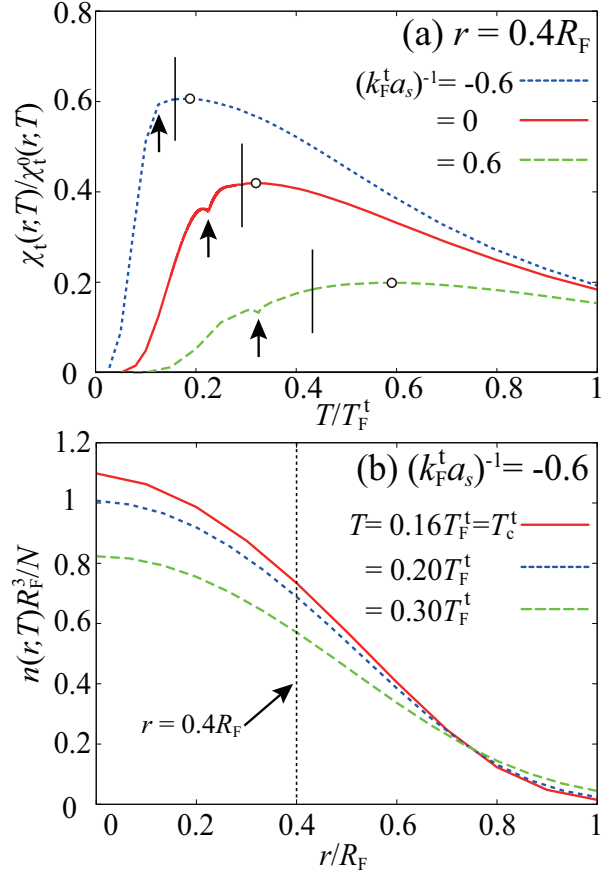}
\end{center}
\caption{(a) Calculated local spin susceptibility $\chi_{\rm t}(r,T)$ at $r=0.4R_{\rm F}$, as a function of temperature. $R_{\rm F}=\sqrt{2\varepsilon_{\rm F}^{\rm t}/(m\Omega_{\rm tr}^2)}$ is the Thomas-Fermi radius, where $\varepsilon_{\rm F}^{\rm t}={k_{\rm F}^{\rm t}}^2/(2m)=(3N)^{1/3}\Omega_{\rm tr}$ is the LDA Fermi energy in a trap (which equals the LDA Fermi temperature $T_{\rm F}^{\rm t}$). $\chi_{\rm t}^0(r,T)=3m n(r,T)^{1/3}/(3\pi^2)^{2/3}$ is the expression for the spin susceptibility in a free Fermi gas at $T=0$ where the number density is replaced by the LDA-ETMA local density $n(r,T)=n_\uparrow(r,T)+n_\downarrow(r,T)$ at $r=0.4R_{\rm F}$. At each line, the short vertical line shows $T^{\rm t}_{\rm c}$, and the open circle represents the peak position of $\chi_{\rm t}(r,T)$ in the normal state. The arrow shows $T^{\rm t}_{\rm c}(r)$, below which the LDA superfluid order parameter $\Delta(r=0.4R_{\rm F})$ becomes non-zero. (b) Density profile $n(r)=n_\uparrow(r)+n_\downarrow(r)$, when $(k^{\rm t}_{\rm F}a_s)^{-1}=-0.6$.
}
\label{fig2}
\end{figure}
\par
\section{Local spin susceptibility and spin-gap phenomenon in a trapped Fermi gas}
\par
Figure \ref{fig2}(a) shows the local spin susceptibility $\chi_{\rm t}(r,T)$ at $r=0.4R_{\rm F}$ (where $R_{\rm F}$ is the Thomas-Fermi radius). In this figure, $\chi_{\rm t}(r,T)$ is found to exhibit a peak structure at a certain temperature ($\equiv T^{\rm t}_{\rm p}(r)$) in the normal state, and is suppressed below this. Since the local superfluid order parameter $\Delta(r)$ only becomes non-zero below the temperature at the arrow in Fig. \ref{fig2}(a), this anomaly is found to occur in the absence of $\Delta(r=0.4R_{\rm F})$.
\par
At a glance, the non-monotonic behavior of $\chi_{\rm t}(r,T)$ around $T_{\rm p}^{\rm t}(r)$ looks similar to the spin-gap phenomenon discussed in the BCS-BEC crossover regime of a {\it uniform} Fermi gas \cite{Tajima}, where this anomaly originates from the formation of spin-singlet preformed (Cooper) pairs. In this magnetic phenomenon, the spin-gap temperature $T_{\rm SG}^{\rm u}$ is defined as the temperature at which the uniform spin susceptibility $\chi_{\rm u}(T)$ takes a maximum value. Regarding this, if the density profile were $T$-independent, each result in Fig. \ref{fig2}(a) could be immediately regarded as the spin susceptibility $\chi_{\rm u}(T)$ in an assumed {\it uniform} Fermi gas with the uniform density $n=n(r=0.4R_{\rm F})=\sum_{\sigma}n_\sigma(r=0.4R_{\rm F})$. However, Fig. \ref{fig2}(b) shows that the density profile $n(r,T)$ actually depends on $T$. Thus, $\chi_{\rm t}(r,T)$ in Fig. \ref{fig2}(a) is also affected by this $T$-dependent density profile, in addition to pairing fluctuations. Since the former effect does not exist in the uniform case, the peak temperature $T_{\rm p}^{\rm t}(r)$ in $\chi_{\rm t}(r,T)$ cannot be immediately identified as the spin-gap temperature $T_{\rm SG}^{\rm u}$ in the uniform case. To examine the spin-gap phenomenon purely originating from pairing fluctuations, we need to remove effects of the $T$-dependent density profile from $\chi_{\rm t}(r,T)$. This would be particularly important, when the local spin susceptibility in a trapped Fermi gas becomes experimentally accessible in the future.
\par
\begin{figure}[t]
\begin{center}
\includegraphics[width=7cm]{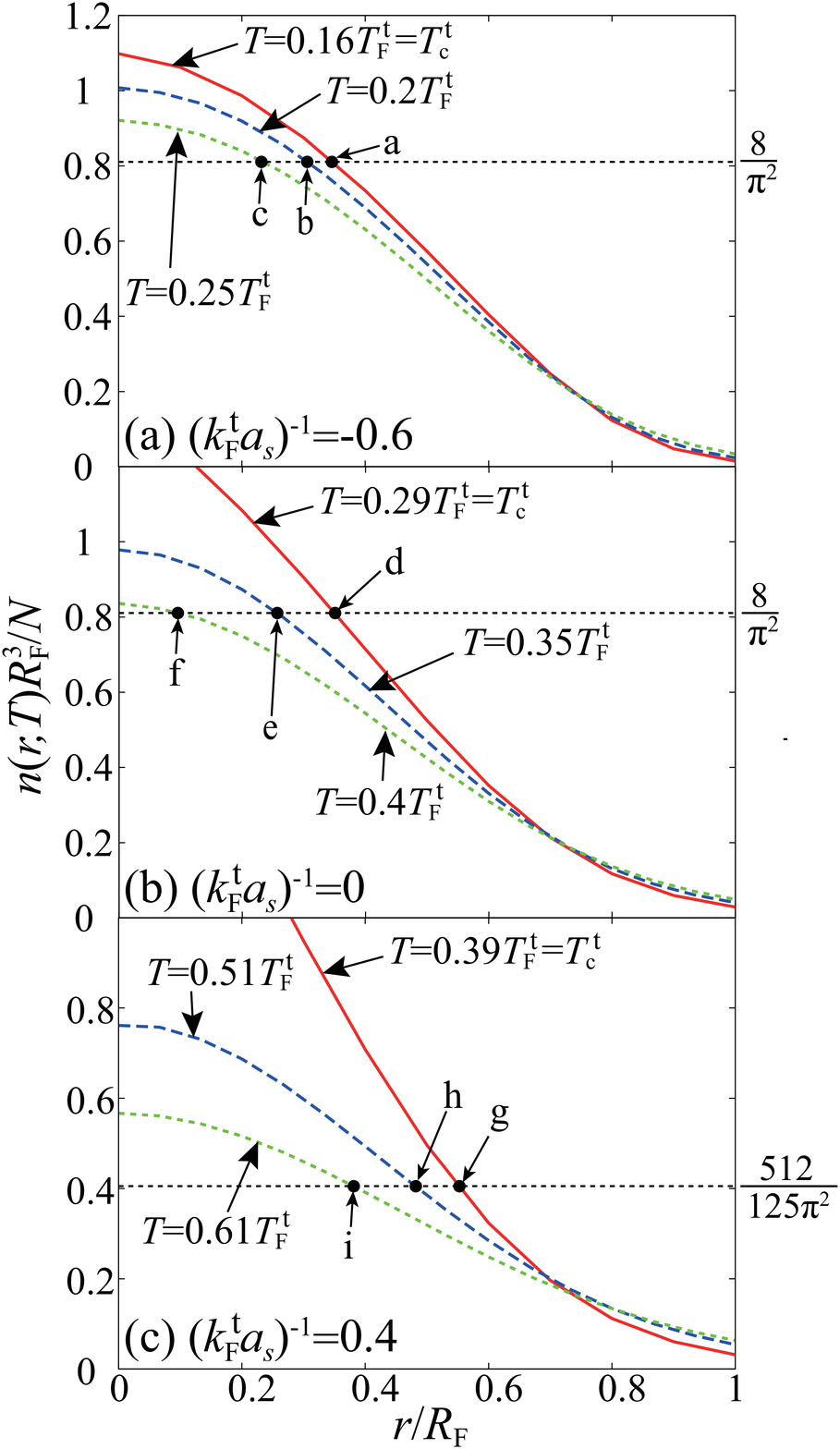}
\end{center}
\caption{Calculated density profile $n(r,T)=\sum_\sigma n_\sigma(r,T)$ in LDA-ETMA. (a) $(k_{\rm F}^{\rm t} a_s)^{-1}=-0.6$. (b) $(k_{\rm F}^{\rm t} a_s)^{-1}=0$ (unitarity limit). (c) $(k_{\rm F}^{\rm t} a_s)^{-1}=0.4$. The filled circles ``a"-``i" correspond to those in Fig.\ref{fig4}(a).}
\label{fig3}
\end{figure}
\par
\begin{figure}[t]
\begin{center}
\includegraphics[width=8cm]{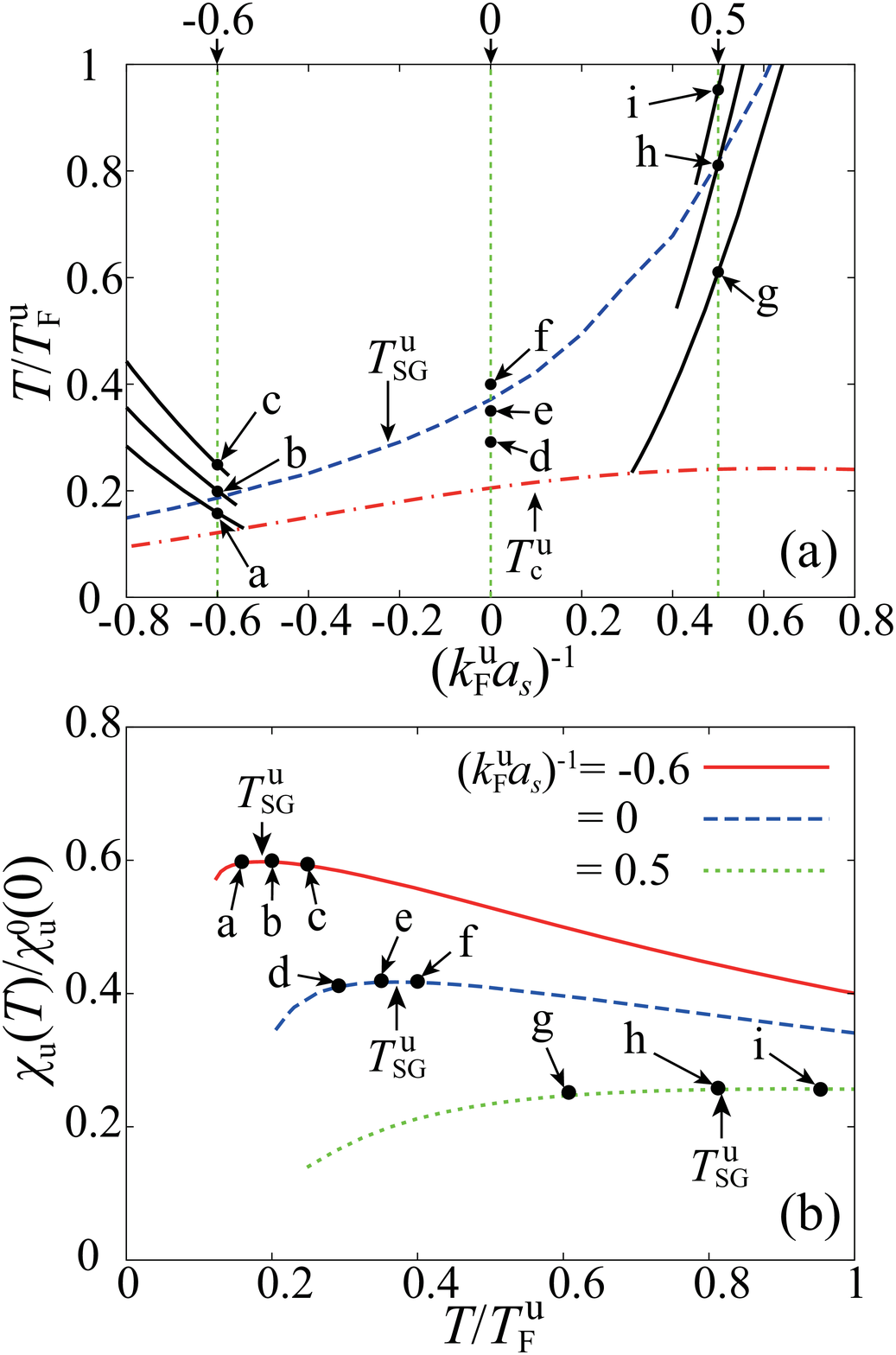}
\end{center}
\caption{Mapping of local spin susceptibility $\chi_{\rm t}(r,T)$ in a trapped Fermi gas onto spin susceptibility $\chi_{\rm u}(T)$ in a uniform Fermi gas. (a) Each solid line with a filled circle (``a"-``i") is the scaled temperature $T/T^{\rm u}_{\rm F}(r)$ as a function of the scaled interaction strength $(k_{\rm F}^{\rm u}a_s)^{-1}$, which is obtained from the local density $n(r,T)$ in Fig. \ref{fig3} at the same label (``a"-``i"). (However, since the solid lines in the cases of ``d"-``f" are the same vertical line at $(k_{\rm F}^{\rm u}a_s)^{-1}=0$, we do not draw them in the figure.) $k_{\rm F}^{\rm u}$, $T_{\rm F}^{\rm u}$, $T_{\rm c}^{\rm u}$, and $T^{\rm u}_{\rm SG}$, are the Fermi momentum, Fermi temperature, the superfluid phase transition temperature, and the spin-gap temperature, in a uniform Fermi gas, respectively. (b) Spin susceptibility $\chi_{\rm u}(T)$ in a uniform Fermi gas \cite{Tajima}. The filled circles ``a"-``i" are the values of the local spin susceptibility $\chi_{\rm t}(r,T)$ at the same labels in Fig. \ref{fig3}. $\chi_{\rm u}^0(0)=mk_{\rm F}^{\rm u}/\pi^2$ is the uniform spin susceptibility in a free Fermi gas at $T=0$.
}
\label{fig4}
\end{figure}
\par
We demonstrate how to extract information about the spin-gap phenomenon in a uniform Fermi gas from the local spin susceptibility $\chi_{\rm t}(r,T)$ in a trapped one. For this purpose, we recall that LDA treats a gas at each spatial position ${\bm r}$ as a {\it uniform} one with the ``(effective) local Fermi momentum",
\begin{equation}
k^{\rm t}_{\rm F}(r,T)=[3\pi^2n(r,T)]^{1/3}.
\label{eq.80aa}
\end{equation}
For example, ``a" in Fig. \ref{fig3}(a) is regarded as a uniform Fermi gas with the Fermi momentum,
\begin{equation}
k^{\rm t}_{\rm F}(r,T)=
\left[
3\pi^2\times{8 \over \pi^2}{N \over R_{\rm F}^3}
\right]^{1/3}=k^{\rm t}_{\rm F}.
\label{eq.80a}
\end{equation}
Here, we have used the LDA relation, $R_{\rm F}=\sqrt{2\varepsilon^{\rm t}_{\rm F}/(m\Omega_{\rm tr}^2)}$, where $\varepsilon^{\rm t}_{\rm F}={k_{\rm F}^{\rm t}}^2/(2m)=(3N)^{1/3}\Omega_{\rm tr}$ is the LDA Fermi energy in a trapped Fermi gas, with $k_{\rm F}^{\rm t}$ being the LDA Fermi momentum \cite{Ohashi6,note80a}. The local spin susceptibility $\chi_{\rm t}(r,T)$ at ``a" in Fig. \ref{fig3}(a) can then be regarded as the susceptibility $\chi_{\rm u}(T)$ in a uniform Fermi gas at the scaled temperature $T/T_{\rm F}^{\rm u}=T/T^{\rm t}_{\rm F}(r)=T/T_{\rm F}^{\rm t}=0.16$, and the scaled interaction strength $(k_{\rm F}^{\rm u} a_s)^{-1}=(k_{\rm F}^{\rm t}(r,T)a_s)^{-1}=(k_{\rm F}^{\rm t}a_s)^{-1}=-0.6$  (``a" in Fig. \ref{fig4}(a)). Here, $T_{\rm F}^{\rm t}={k_{\rm F}^{\rm t}}^2/(2m)$ is the LDA Fermi temperature in a {\it trapped} Fermi gas, and $k_{\rm F}^{\rm u}$ and $T_{\rm F}^{\rm u}={k_{\rm F}^{\rm u}}^2/(2m)$ are the Fermi momentum and Fermi temperature in a {\it uniform} Fermi gas, respectively. In the same manner, the spatial position ``b" and ``c" in Fig. \ref{fig3}(a) are mapped onto the uniform system with the same scaled interaction strength $(k_{\rm F}^{\rm u} a_s)^{-1}=-0.6$, but at $T/T_{\rm F}^{\rm u}=0.2$ and 0.25, respectively (see Fig. \ref{fig4}(a)). As shown in Fig. \ref{fig4}(b), the values of $\chi_{\rm t}(r,T)$ at ``a"-``c" in Fig. \ref{fig3}(a) coincide with the previous ETMA result for a uniform Fermi gas at $(k_{\rm F}^{\rm u}a_s)^{-1}=-0.6$ \cite{Tajima}, as expected.
\par
The above prescription is also valid for stronger coupling cases. Indeed, the positions ``d"-``i" in Figs. \ref{fig3}(b) and (c) are mapped onto the uniform case at the same labels in Fig. \ref{fig4}, respectively. 
\par
We note that this mapping can be simplified to some extent at the unitarity, because $\chi_{\rm t}(r,T)$ in this special case is always mapped onto $\chi_{\rm u}(T)$ in a uniform unitary Fermi gas. (Note that the scaled interaction $(k_{\rm F}^{\rm u}a_s)^{-1}$ identically vanishes when $a_s^{-1}=0$, irrespective of the value of the Fermi momentum $k_{\rm F}^{\rm u}$.) Using this, we can construct the temperature dependence of $\chi_{\rm u}(T)$ at the unitarity only from the temperature dependence of $\chi_{\rm t}(r,T)$ at a fixed position $r$. The maximum $\chi_{\rm t}(r,T)$ is mapped onto the maximum $\chi_{\rm u}(T)$ in this case, so that one can exceptionally relate the peak temperature $T_{\rm p}^{\rm t}(r)$ in the trapped case (open circle in Fig. \ref{fig2}(a)) to the spin-gap temperature $T_{\rm SG}^{\rm u}$ in the uniform case as,
\begin{equation}
{T_{\rm SG}^{\rm u} \over T_{\rm F}^{\rm u}}=
\left(
{8 \over \alpha(r)\pi^2}
\right)^{2/3}
{T_{\rm p}^{\rm t}(r) \over T_{\rm F}^{\rm t}},
\label{eq.80b}
\end{equation}
where $\alpha(r)=(R_{\rm F}^3/N)n(r,T_{\rm p}^{\rm t}(r))$.
\par
\begin{figure}[t]
\begin{center}
\includegraphics[width=7cm]{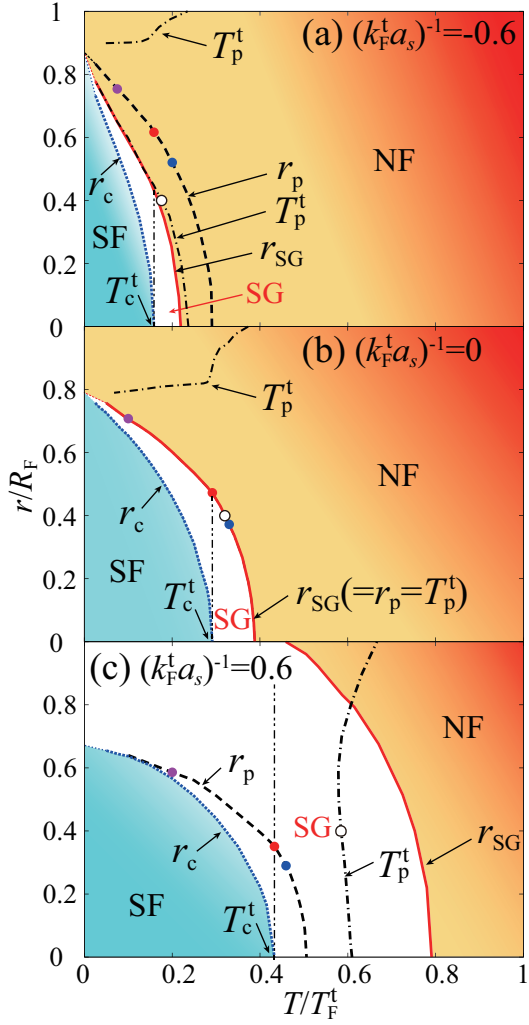}
\end{center}
\caption{$r-T$ phase diagram of a trapped Fermi gas. (a) $(k^{\rm t}_{\rm F}a_s)^{-1}=-0.6$ (weak-coupling BCS side). (b) $(k^{\rm t}_{\rm F}a_s)^{-1}=0$ (unitarity limit). (c) $(k^{\rm t}_{\rm F}a_s)^{-1}=0.6$ (strong-coupling BEC side). The LDA superfluid order parameter $\Delta(r)$ becomes non-zero when $r\le r_{\rm c}(T)$ (SF). $T_{\rm p}^{\rm t}(r)$ is the temperature at which $\chi_{\rm t}(r,T)$ takes a maximum value, when $r$ is fixed. $r_{\rm p}(T)$ is the spatial position at which $\chi_{\rm t}(r,T)$ takes a maximum value, when $T$ is fixed. $\chi_{\rm t}(r_{\rm SG}(T),T)$ is mapped onto $\chi_{\rm u}(T_{\rm SG}^{\rm u})$ in a uniform Fermi gas. The region $r_{\rm c}(T)\le r\le r_{\rm SG}(T)$ (SG) is mapped onto the spin-gap regime in a uniform Fermi gas, where $\chi_{\rm u}(T)$ is suppressed by pairing fluctuations. The region $r>r_{\rm SG}(T)$ (NF) is mapped onto the normal Fermi gas regime in the uniform case, where $\chi_{\rm u}(T)$ monotonically increases with decreasing the temperature. The open and filled circles represent $T_{\rm p}^{\rm t}(r)$ and $r_{\rm p}(T)$ obtained from Figs. \ref{fig2} and \ref{fig6}, respectively. Because of computational problems at low temperatures ($T\lesssim 0.02T^{\rm t}_{\rm F}$), we only draw eye-guide (thin dashed line) for each line there.
}
\label{fig5}
\end{figure}
\par
Figure \ref{fig5} shows the phase diagram of a trapped Fermi gas with respect to the spatial position $r$ (measured from the trap center) and the temperature $T$ in LDA-ETMA. In each panel, $r_{\rm SG}(T)$ is the spatial position which is mapped onto the spin-gap temperature $T_{\rm SG}^{\rm u}$ in a uniform Fermi gas with the uniform density $n=n(r_{\rm SG}(T),T)$ and the interaction strength $(k_{\rm F}^{\rm u}a_s)^{-1}=(k_{\rm F}^{\rm t}(r_{\rm SG}(T))a_s)^{-1}$, for a given interaction strength $(k_{\rm F}^{\rm t}a_s)^{-1}$. As expected, one sees in Fig. \ref{fig5}(b) that the peak temperature $T_{\rm p}^{\rm t}(r)$ coincides with the ``spin-gap line" $r_{\rm SG}(T)$ in the unitarity limit, except for the outer region of the gas cloud, $r\gesim 0.8R_{\rm F}$ (which will be separately discussed later). 
\par
Although this coincidence is only guaranteed at the unitarity, Fig. \ref{fig5}(a) shows that $T_{\rm p}^{\rm t}(r)$ is still close to $r_{\rm SG}(T)$ in the weak-coupling BCS side (as far as we consider the region $r\lesssim 0.8R_{\rm F}$). Thus, the peak-temperature $T_{\rm p}^{\rm t}(r)$ in the trapped case is still useful for {\it roughly} estimating the spin-gap temperature $T_{\rm SG}^{\rm u}$ in the BCS side of a uniform Fermi gas. On the other hand, we see in Fig. \ref{fig5}(c) that $T_{\rm p}^{\rm t}(r)$ is very different from $r_{\rm SG}(T)$ in the BEC side, indicating that we need to faithfully fulfil the above-mentioned mapping, in order to examine the spin-gap there. 
\par
The LDA superfluid order parameter $\Delta(r)$ only becomes non-zero when $T\le T_{\rm c}^{\rm t}(r)~(\le T_{\rm c}^{\rm t})$, which leads to the shell structure of the system below $T_{\rm c}^{\rm t}$, being composed of the superfluid core region ($\Delta(r\le r_{\rm c}(T))\ne 0$) which is surrounded by the normal-fluid region ($\Delta(r>r_{\rm c}(T))=0$). In this case, the region ``SG" in Fig. \ref{fig5} ($r_{\rm c}(T)\le r\le r_{\rm SG}(T)$) is mapped onto the spin-gap regime ($T_{\rm c}^{\rm u} \le T\le T_{\rm SG}^{\rm u}$) of an uniform Fermi gas, where $\chi_{\rm u}(T)$ is suppressed by pairing fluctuations (where $T_{\rm c}^{\rm u}$ is the superfluid phase transition temperature in the uniform case). The region ``NF" and ``SF" in Fig. \ref{fig5}, are, respectively, mapped onto the normal Fermi gas regime (where $\chi_{\rm u}(T)$ monotonically increases as the temperature decreases), and the superfluid regime (where $\chi_{\rm u}(T)$ is suppressed by the superfluid order) of a uniform Fermi gas, respectively.
\par
\begin{figure}[t]
\begin{center}
\includegraphics[width=7cm]{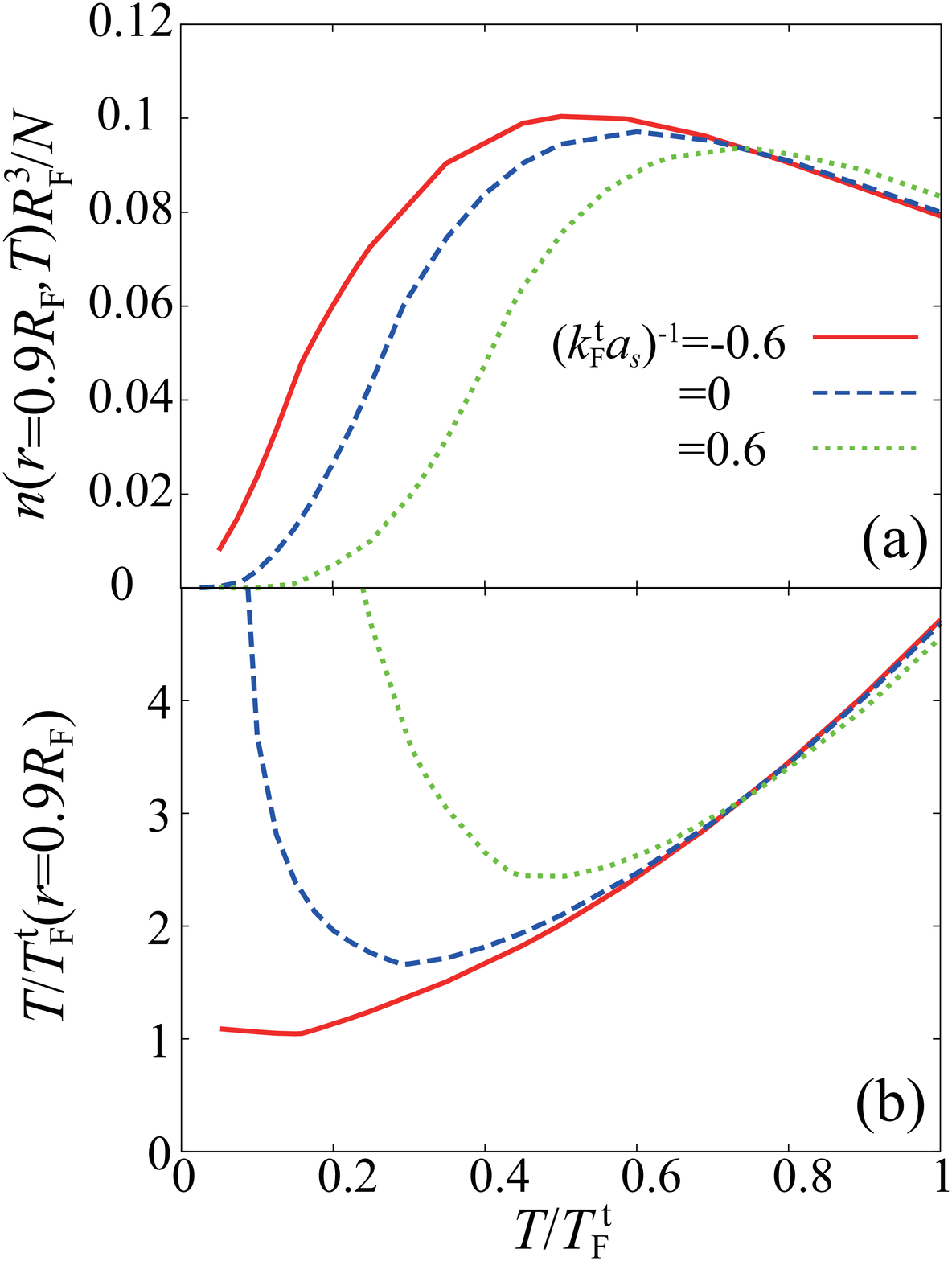}
\end{center}
\caption{(a) Temperature dependence of local density $n(r,T)$ in the outer region of the gas cloud at $r=0.9R_{\rm F}$. (b) Scaled local temperature $T/T_{\rm F}^{\rm t}(r=0.9R_{\rm F})$, as a function of $T/T_{\rm F}^{\rm t}$. 
}
\label{fig6}
\end{figure}
\par
Of course, the above-mentioned shell structure is, strictly speaking, an artifact of LDA. The superfluid order parameter $\Delta(r)$ should actually become non-zero everywhere in a gas below $T_{\rm c}^{\rm t}$. Thus, when we experimentally examine the spin-gap phenomenon purely caused by {\it normal-state} pairing fluctuations, we should examine the region surrounded by the vertical $T_{\rm c}^{\rm t}$-line and the spin-gap line $r_{\rm SG}(T)$ in Fig. \ref{fig5}.
\par
As briefly mentioned previously, in the unitarity limit shown in Fig. \ref{fig5}(b), while the peak temperature $T_{\rm p}^{\rm t}(r)$ coincides with the spin-gap line $r_{\rm SG}(T)$ in the central region of the gas cloud ($r\lesssim 0.8R_{\rm F}$), such coincidence is not obtained in the outer region, $r\gesim 0.8R_{\rm F}$, implying that $T_{\rm p}^{\rm t}(r\gesim 0.8R_{\rm F})$ comes from a different origin from the spin-gap phenomenon. To understand the origin of this peak temperature, the key is that, when one increases the temperature from $T=0$, the local density $n(r\gesim 0.8R_{\rm F})$ first increases because of the thermal expansion of the gas cloud, as shown in Fig. \ref{fig6}(a). As a result, the scaled local temperature $T/T_{\rm F}^{\rm t}(r\gesim 0.8R_{\rm F})$ exhibits a non-monotonic temperature dependence, as shown in Fig. \ref{fig6}(b). In the case of Fig. \ref{fig6} ($r=0.9R_{\rm F}$), denoting the dip temperature in Fig. \ref{fig6}(b) as $T_{\rm dip}$, one finds that the increase of $T/T_{\rm F}^{\rm t}$ in the low temperature region of a trapped Fermi gas ($T\le T_{\rm dip}$) corresponds to the {\it decrease} of $T/T^{\rm u}_{\rm F}=T/T_{\rm F}^{\rm t}(r=0.9R_{\rm F})$ in the high-temperature region of a {\it uniform} Fermi gas. Thus, reflecting the increasing of $\chi_{\rm u}(T)$ with decreasing $T/T^{\rm u}_{\rm F}$ in the high-temperature region, the corresponding $\chi_{\rm t}(r=0.9R_{\rm F},T)$ {\it increases} with increasing $T/T_{\rm F}^{\rm t}$ when $T\le T_{\rm dip}$. On the other hand, when $T\ge T_{\rm dip}$, the increase of $T/T_{\rm F}^{\rm t}$ corresponds to the {\it increase} of $T/T^{\rm u}_{\rm F}$. Thus, the decrease of $\chi_{\rm u}(T)$ with increasing $T/T_{\rm F}^{\rm t}$ leads to the {\it decrease} of $\chi_{\rm t}(r=0.9R_{\rm F},T)$ with increasing $T/T_{\rm F}^{\rm t}$ when $T\ge T_{\rm dip}$.
\par
To conclude, although the resulting $\chi_{\rm t}(r\gesim 0.8R_{\rm F})$ takes a maximum value at $T_{\rm dip}$, it is clearly not due to pairing fluctuations, but simply originates from the temperature dependence of the density profile around the edge of the gas cloud. Since the non-monotonic behavior of $T/T_{\rm F}^{\rm t}(r=0.9R_{\rm F})$ is also seen in the other two cases shown in Fig. \ref{fig6}, $T_{\rm p}^{\rm t}(r\gesim 0.8R_{\rm F})$ in Fig. \ref{fig5}(a), as well as that in Fig. \ref{fig5}(c), are also nothing to do with the spin-gap phenomenon.
\par
Regarding the above-mentioned effects of $T$-dependent density profile, we briefly note that, while the thermal expansion of the trapped gas increases the density $n(r,T)$ in the outer region of the gas cloud at low temperatures, it decreases $n(r)$ in the central region, as seen in Fig. \ref{fig3}. Because of this, the scaled local temperature $T/T_{\rm F}^{\rm t}(r)$ in the trap center monotonically increases with increasing $T/T_{\rm F}^{\rm t}$. Thus, the increase of $T/T_{\rm F}^{\rm t}$ in the trapped case can simply be related to the increase of $T/T_{\rm F}^{\rm u}$ in the uniform case there. 
\par
\begin{figure}[t]
\begin{center}
\includegraphics[width=8cm]{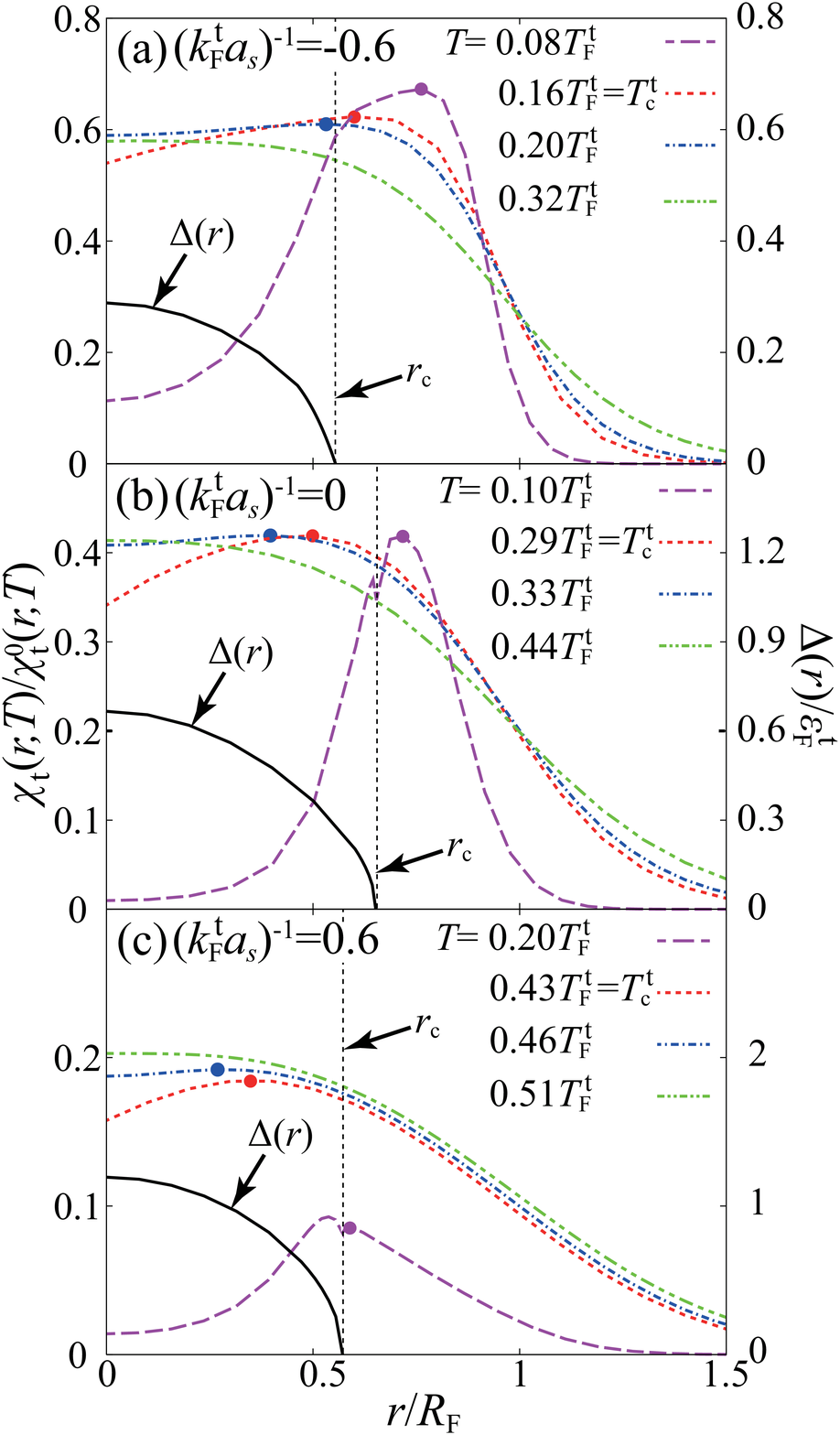}
\end{center}
\caption{Local spin susceptibility $\chi_{\rm t}(r,T)$, as a function of the spatial position $r$, measured from the trap center. We also plot the superfluid order parameter $\Delta(r)$.
}
\label{fig7}
\end{figure}
\par
Figure \ref{fig7} shows the spatial variation of $\chi_{\rm t}(r,T)$ in a trapped Fermi gas. In addition to the well-known suppression of spin susceptibility in the superfluid phase ($\Delta(r,T)\ne 0$), $\chi_{\rm t}(r,T)$ is found to be suppressed in the trap center ($r\sim 0$), even in the normal state. Conveniently defining the peak radius $r_{\rm p}(T)$ as the position at which the spatial variation of $\chi_{\rm t}(r,T)$ takes a maximum value above $T_{\rm c}^{\rm t}$, we find that it agrees with the spin-gap radius $r_{\rm SG}(T)$ in the unitarity limit (see Fig. \ref{fig6}(b)). This is simply because the scaled local interaction strength $(k_{\rm F}^{\rm t}(r)a_s)^{-1}$ always vanishes at the unitarity ($a_s^{-1}=0$), irrespective of the value of $k_{\rm F}^{\rm t}(r)$, so that $\chi_{\rm t}(r,T)$ in the unitarity limit is always mapped onto $\chi_{\rm u}(T)$ in a uniform unitary Fermi gas. This means that we can evaluate the spin-gap temperature without measuring the temperature dependence of $\chi_{\rm t}(r,T)$ at the unitarity. 
\par
Of course, the peak radius $r_{\rm p}(T)$ does not coincide with the spin-gap line $r_{\rm SG}(T)$ for $(k_{\rm F}^{\rm t}a_s)^{-1}\ne 0$ (see Figs. \ref{fig5}(a) and (c)), because of the position dependent $T/T_{\rm F}^{\rm t}(r)$, and $(k_{\rm F}^{\rm t}(r)a_s)^{-1}$.
\par
\begin{figure}[t]
\begin{center}
\includegraphics[width=8cm]{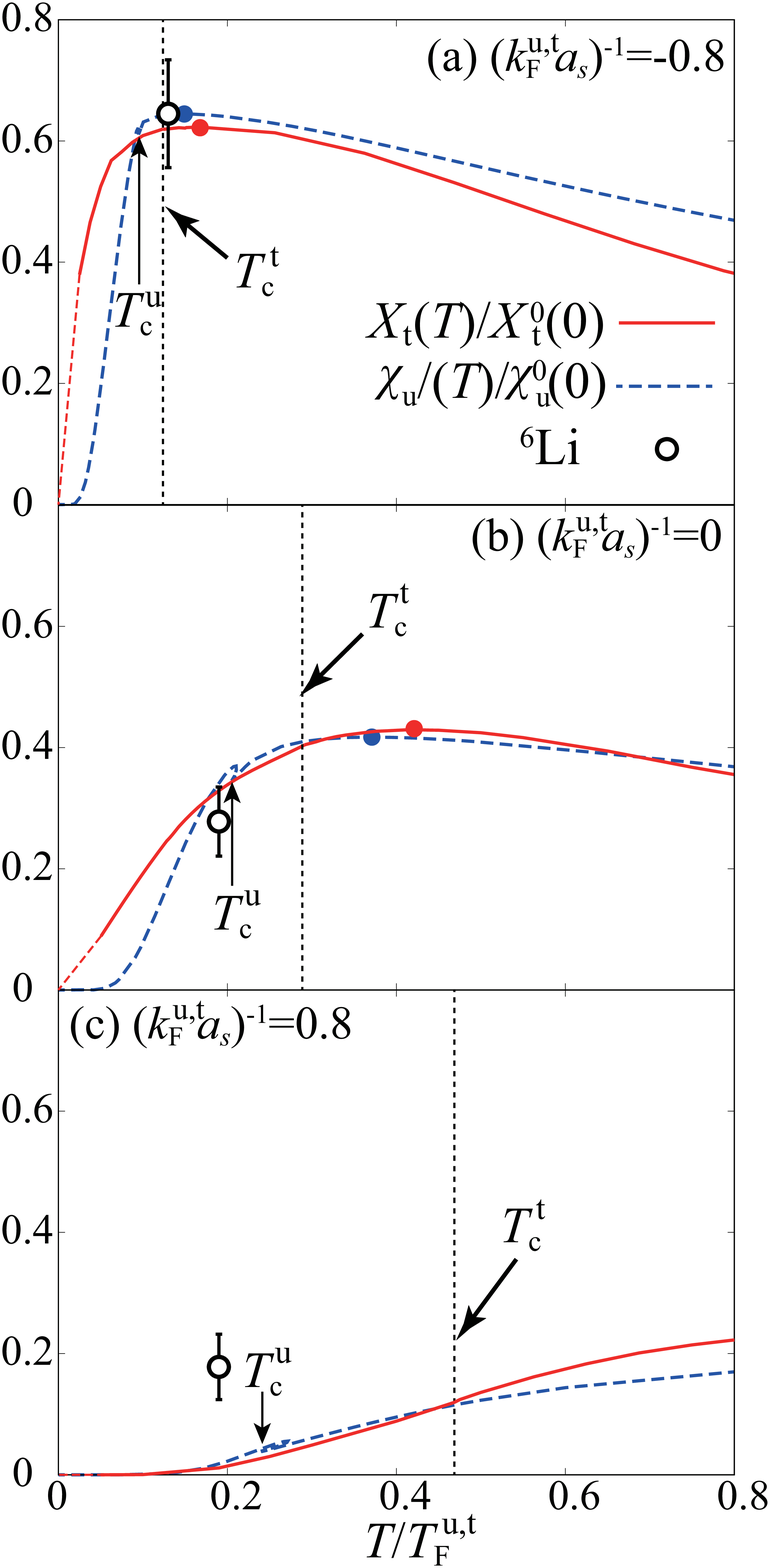}
\end{center}
\caption{Calculated trap-averaged spin susceptibility $X_{\rm t}(T)$ in Eq. (\ref{eq12}), normalized by the value $X_{\rm t}^0(0)=3N/\varepsilon_{\rm F}^{\rm t}$ in a trapped free Fermi gas at $T=0$. (a) $(k^{\rm t}_{\rm F}a_s)^{-1}=-0.8$. (b) $(k^{\rm t}_{\rm F}a_s)^{-1}=0$. (c) $(k^{\rm t}_{\rm F}a_s)^{-1}=-0.8$. For comparison, we also plot the ETMA spin susceptibility $\chi_{\rm u}(T)$ in a uniform Fermi gas\cite{Tajima2}, normalized by the value $\chi_{\rm u}^0(0)=mk_{\rm F}^{\rm u}/\pi^2$ in a free Fermi gas at $T=0$. In each result, the filled circle shows the temperature at which the spin susceptibility takes a maximum value. In the uniform case, it gives the spin-gap temperature $T_{\rm SG}^{\rm u}$. In the trapped case, it gives ${\tilde T}_{\rm p}^{\rm t}$. The open circles are the recent experimental data on a $^6$Li Fermi gas \cite{Sanner}. Because of computational problems, our LDA-ETMA results end at $T\simeq 0.02T^{\rm t}_{\rm F}$; the thin dashes lines at lower temperatures in panels (a) and (b) are eye-guide.
}
\label{fig8}
\end{figure}
\par
\section{Trap-averaged spin susceptibility in the BCS-BEC crossover region}
\par
Figure \ref{fig8} compares the trap-averaged spin susceptibility $X(T)$ in Eq. (\ref{eq12}) with the spin susceptibility $\chi_{\rm u}(T)$ in a uniform Fermi gas. We find that the behavior of $X(T)$ is relatively close to that of $\chi_{\rm u}(T)$, in spite of the fact that $X_{\rm t}(T)$ is affected by $T$-dependent density profile $n(r,T)$. Figure \ref{fig8} also shows that the both $X(T)$ and $\chi_{\rm u}(T)$ agree with the recent experiment on a $^6$Li Fermi gas\cite{Sanner} in the weak-coupling regime, as well as in the unitarity limit. Although the spatial resolution of this experiment\cite{Sanner} is unclear, our results indicate that the spatial inhomogeneity is not so crucial for the observed spin susceptibility, at least in the cases of Figs. \ref{fig8}(a) and (b). 
\par
In our previous paper \cite{Tajima2}, we pointed out the the observed spin susceptibility in the strong-coupling BEC side ($(k_{\rm F}^{\rm u}a_s)^{-1}=0.8>0$) cannot be explained by ETMA spin susceptibility in a uniform Fermi gas. In this regard, Fig. \ref{fig8}(c) shows that this problem still remains in the trapped case, because $X_{\rm t}(T)$ is still much smaller than the observed value. In order to reproduce the experimental result in the strong-coupling regime \cite{Sanner} within the current LDA-ETMA formalism, we need to raise the temperature to $T\simeq 0.6T_{\rm F}^{\rm t}$. At preset, we have no idea to fill up this discrepancy, which remains our future problem.
\par
\begin{figure}[t]
\begin{center}
\includegraphics[width=8cm]{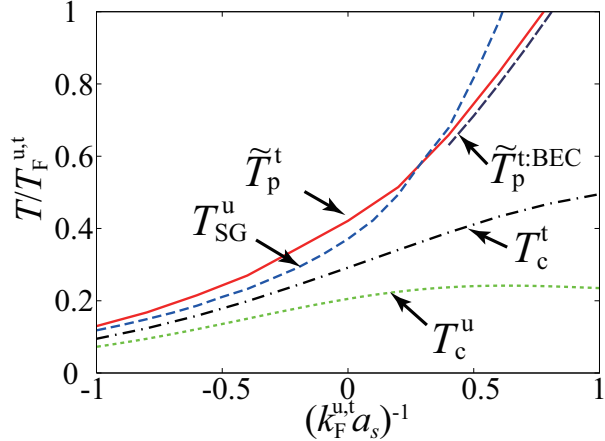}
\end{center}
\caption{Peak temperature ${\tilde T}_{\rm p}^{\rm t}$ of $X_{\rm t}(T)$. For comparison, we also plot the spin-gap temperature $T_{\rm SG}^{\rm u}$ \cite{Tajima}. ${\tilde T}_{\rm p}^{\rm t:BEC}$ is the solution of Eq. (\ref{eq30}).
}
\label{fig9}
\end{figure}
\par
\begin{figure}[t]
\begin{center}
\includegraphics[width=8cm]{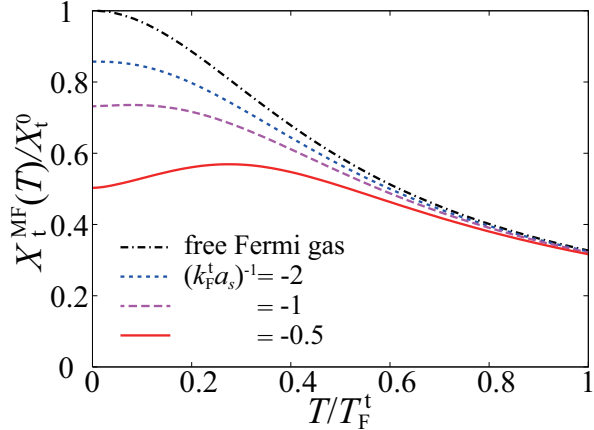}
\end{center}
\caption{Averaged spin susceptibility $X_{\rm t}^{\rm MF}(T)$ in the mean-field approximation, where the Hartree-Fock mean-field self-energy $\hat{\Sigma}^{\rm MF}(r,T)$ in Eq. (\ref{eq5HF}) is used for the ETMA one in Eq. (\ref{eq5}).}
\label{fig10}
\end{figure}

Figure \ref{fig9} shows the peak temperature ${\tilde T}_{\rm p}^{\rm t}$ at which the averaged spin susceptibility $X_{\rm t}(T)$ takes a maximum value. As expected from the similarity between $X_{\rm t}(T)$ and $\chi_{\rm u}(T)$ in Fig. \ref{fig8}, ${\tilde T}_{\rm p}^{\rm t}$ is relatively close to the spin-gap temperature $T_{\rm SG}^{\rm u}$ in a uniform Fermi gas, although the former also involves effects of the $T$-dependent density profile. Indeed, when we ignore pairing fluctuations by replacing the ETMA self-energy in Eq. (\ref{eq5}) with that in the mean-field approximation \cite{FW,noteMF},
\par
\begin{equation}
\hat{\Sigma}^{\rm MF}(r,T)=\frac{4\pi a_s}{m}\left[n_{\dwn}(r,T)\frac{(1+\tau_3)}{4}-n_{\up}(r,T)\frac{(1-\tau_3)}{4}\right],
\label{eq5HF}
\end{equation}
the resulting averaged spin susceptibility ($\equiv X_{\rm t}^{\rm MF}(T)$) exhibits ``spin-gap" like temperature dependence, as shown in Fig. \ref{fig10}. Since the averaged spin susceptibility does not exhibit such a non-monotonic behavior when the density profile is $T$-independent, it purely comes from the $T$-dependent $n_\sigma(r,T)$. The peak temperature ${\tilde T}_{\rm p}^{\rm t}$ is considered to also involve this effect, in addition to spin-gap effects associated with pairing fluctuations.
\par
\par
In Ref. \cite{Tajima}, we showed that the spin-gap temperature $T_{\rm SG}^{\rm u}$ in the strong coupling regime of a uniform Fermi gas can be explained by a classical gas mixture, consisting of two kinds of atoms with active spins $\sigma=\uparrow,\downarrow$ and one-component spinless molecules \cite{Saha}. When we simply extend this to the present trapped case, the equation for the peak temperature ${\tilde T}_{\rm p}^{\rm t:BEC}$ of $X_{\rm t}(T)$ in this classical gas mixture is obtained as,
\begin{equation}
{(2mR_{\rm F}^2{\tilde T}_{\rm p}^{\rm t:BEC})^3 \over 108N^2
}\exp \left(-\frac{E_{\rm b}}{T}\right)=
\frac{\left[\left(\frac{E_{\rm b}+3{\tilde T}_{\rm p}^{\rm t:BEC}}{E_{\rm b}+2{\tilde T}_{\rm p}^{\rm t:BEC}}\right)-2\right]^2}{\frac{E_{\rm b}+3{\tilde T}_{\rm p}^{\rm t:BEC}}{E_{\rm b}+2{\tilde T}_{\rm p}^{\rm t:BEC}}-1},
\label{eq30}
\end{equation}
where $E_{\rm b}=1/(ma_s^2)$ is the molecular binding energy. (For the derivation of Eq. (\ref{eq30}), see the Appendix.) The calculated ${\tilde T}_{\rm p}^{\rm t:BEC}$ well reproduces ${\tilde T}_{\rm p}^{\rm t}$ in the strong-coupling regime (see Fig. \ref{fig10}), indicating that the simple classical gas mixture is also valid in considering $X_{\rm t}(T)$ in a trapped Fermi gas when $(k_{\rm F}^{\rm t}a_s)\gesim 0.5$.
\par
\par
\section{Summary}
\par
To summarize, we have discussed magnetic properties of a trapped ultracold Fermi gas. Including effects of strong pairing fluctuations within in the framework of an extended $T$-matrix approximation (ETMA), as well as effects of a harmonic trap in the local density approximation (LDA), we have calculated local spin susceptibility $\chi_{\rm t}(r,T)$, as well as the spatially averaged one in the whole BCS-BEC crossover region. 
\par
We showed that the local spin susceptibility $\chi_{\rm t}(r,T)$ in the BCS-BEC crossover region exhibits a non-monotonic temperature dependence, taking a maximum value at a certain temperature $T^{\rm t}_{\rm p}(r)$. At a glance, it looks similar to the spin-gap behavior of the spin susceptibility $\chi_{\rm u}(T)$ in a uniform Fermi gas. However, the former peak temperature $T^{\rm t}_{\rm p}(r)$ cannot actually be simply related to the latter spin-gap temperature $T^{\rm u}_{\rm SG}$ (except at the unitarity), because the former also involves effects of {\it temperature-dependent} density profile, in addition to effects of pairing fluctuations. We explained how to evaluate $T^{\rm u}_{\rm SG}$, by properly mapping $\chi_{\rm t}(r,T)$ onto $\chi_{\rm u}(T)$. Using this, we also identified the region which is mapped onto the spin-gap regime ($T_{\rm c}^{\rm u}\le T\le T^{\rm u}_{\rm SG}$) of a uniform Fermi gas, in the phase digram of a trapped Fermi gas with respect to the spatial position $r$ measured from the trap center and the temperature. 
\par
We pointed out that this mapping can be simplified to some extent in the unitarity limit, because the local spin susceptibility $\chi_{\rm t}(r,T)$ in a trapped {\it unitary} Fermi gas is always mapped onto $\chi_{\rm u}(T)$ in a uniform {\it unitary} Fermi gas. Using this advantage, we can immediately relate the peak temperature $T_{\rm p}^{\rm t}$ to the spin-gap temperature $T^{\rm u}_{\rm SG}$, by way of the simple relation in Eq. (\ref{eq.80b}). We pointed out that this advantage also enables us to evaluate $T_{\rm SG}^{\rm u}$ from the spatial variation of $\chi_{\rm t}(r,T)$ for a fixed temperature.
\par
Besides the local spin susceptibility, we also examined the spatially averaged spin susceptibility $X_{\rm t}(T)$. The calculated $X_{\rm t}(T)$ was shown to agree with the recent experiment on a $^6$Li Fermi gas in the weak-coupling regime, as well as in the unitarity limit. However, in the strong-coupling BEC regime, our result was found to be much smaller than the observed value. In this regard, our previous work for a uniform Fermi gas has already faced the same discrepancy in the strong-coupling regime \cite{Tajima}. Thus, our result in this paper indicates that this problem is nothing to do with effects of a harmonic trap. Explaining theoretically the observed large spin susceptibility in the strong-coupling regime remains as our future problem.
\par
Even when the local measurement of spin susceptibility in an ultracold Fermi gas becomes possible in the future, experimental data would more or less involve effects of finite spatial resolution. In this regard, this paper has only dealt with the two extreme cases, that is, the local spin susceptibility $\chi_{\rm t}(r,T)$ and the fully averaged one $X_{\rm t}(T)$. Thus, as a future challenge, it would be interesting to theoretically clarify the minimal spatial resolution which is necessary to examine the spin-gap phenomenon, by using the observed spin susceptibility in a trapped Fermi gas. We briefly note that this kind of theoretical estimation has recently been done \cite{Ota} for the local photoemission-type experiment developed by JILA group \cite{Sagi2}. At present, because cold atom physics has no experimental technique to directly observe the pseudogapped density of states, our results would be useful for the assessment of preformed pair scenario from the viewpoint of spin-gap phenomenon in a trapped ultracold Fermi gas.
\par
\acknowledgements
\par
We thank T. Kashimura, R. Watanabe, D. Inotani, and M. Ota for useful discussions. This work was supported by KiPAS project at Keio University. H.T. and R.H. were supported by a Grant-in-Aid for JSPS fellows (No.17J03975, No.17J01238). Y.O. was also supported by Grant-in-Aid for Scientific research from MEXT and JSPS in Japan (No.16K05503, No.15K00178, No.15H00840).
\par
\appendix
\section{Derivation of Eq. (\ref{eq30})}
\label{apA}
We consider a non-interacting classical gas mixture, consisting of two-component atoms with active spins $\sigma=\uparrow,\downarrow$ (with the number density $n^0_\sigma(r,T)$) and one-component spinless molecules (with the molecular density $n_{\rm M}(r,T)$), in a harmonic trap potential. In the BEC regime of an ultracold Fermi gas, the former two and the latter correspond to unpaired Fermi atoms and tightly bound molecular bosons, respectively. The total atomic number density $n(r,T)$ is given by
\begin{equation}
n(r,T)=n^0_\uparrow(r,T)+n^0_\downarrow(r,T)+2n_{\rm M}(r,T),
\label{eq25}
\end{equation}
where
\begin{eqnarray}
\label{eq26}
n^0_\sigma(r,T)
=\sum_{\bm{p}}\exp\left[-\frac{\xi_{\bm{p}}(r)-\s h}{T}\right]
=\frac{3\sqrt{\pi}}{8}\left(\frac{T}{T_{\rm F}^{\rm t}}\right)^{\frac{3}{2}}
\lambda\exp\left(\frac{\s h - m\Omega_{\rm tr}^2r^2/2}{T}\right),
\end{eqnarray}   
\begin{eqnarray}
\label{eq27}
n_{\rm M}(r)=\sum_{\bm{q}}\exp\left[-\frac{\varepsilon_{\bm{q}}^{\rm M}-2\mu(r)-E_{\rm b}}{T}\right]
=\frac{3\sqrt{2\pi}}{4}\lambda^2
\exp\left(\frac{E_{\rm b}-m\Omega_{\rm tr}^2r^2}{T}\right).
\end{eqnarray} 
Here, $\varepsilon_{\bm q }^{\rm M}=\bm{q}^2/(4m)$ is the molecular kinetic energy and $\lambda=\exp(\mu/T)$ is the fugacity. Solving the total number equation,\begin{equation}
N=\int d{\bm r}n(r,T),
\label{eq27a}
\end{equation}
in terms of the fugacity $\lambda$, one obtains,
\begin{equation}
\label{eq28}
\lambda=
{1 \over 2}
\exp\left(-\frac{E_{\rm b}}{T}\right)
\left[\sqrt{1+{2 \over 3}\left(\frac{T^{\rm t}_{\rm F}}{T}\right)^3
\exp\left(\frac{E_{\rm b}}{T}\right)}-1\right].
\end{equation}
\par
Noting that the averaged spin susceptibility ($\equiv X_{\rm t}^{\rm cl}(T)$) in the present model classical gas is obtained from spin-active atoms.
Therefore, we reach,
\begin{eqnarray}
\label{eq29}
X_{\rm t}^{\rm cl}(T)=
\lim_{h\rightarrow 0}\int d\bm{r}
{n_{\up}^0(r)-n_{\dwn}^0(r) \over h}
= 2\left({T \over T^{\rm t}_{\rm F}}\right)^2\lambda X_{\rm t}^0(0).
\end{eqnarray}
Equation (\ref{eq30}) is straightforwardly obtained from the extremum condition $(\partial X_{\rm t}^{\rm cl}/\partial T)=0$.
\par
\par


\begin{thebibliography}{99}
\bibitem{note80a} In LDA, the Fermi momentum $k^{\rm t}_{\rm F}$ is related to the central density $n_0(r=0)$ of a trapped free Fermi gas at $T=0$ as $k^{\rm t}_{\rm F}=[3\pi^2n_0(r=0)]^{1/3}$. Noting that the density profile in a free Fermi gas is  given by $n_0(r)=n_0(r)[1-(r/R_{\rm F})^2]$, we can rewrite this relation into $k_{\rm F}^{\rm t}=(12N)^{1/3}\Omega_{\rm tr}$, for a given total number $N$ of Fermi atoms in a trap.
\bibitem{noteMF} In evaluating Eq. (\ref{eq5HF}), the local density profile $n(r,T)$ has been self-consistently determined within the framework of the combined mean-field approximation with LDA.
\end{thebibliography}
\end{document}